\documentclass[aps,prd,twocolumn,superscriptaddress,showpacs,amsmath,amssymb,nofootinbib, preprintnumbers]{revtex4-2}
 \usepackage{amsmath}
\usepackage{graphicx}
\usepackage{epstopdf}
\usepackage{float}
\usepackage{hyperref}
\usepackage{color}
\usepackage[T1]{fontenc}
\usepackage[utf8]{inputenc}
\usepackage[toc,page]{appendix}
\usepackage[usenames,dvipsnames]{xcolor}
\usepackage[normalem]{ulem}

\newcommand{\be}{\begin{equation}}
\newcommand{\ee}{\end{equation}}
\newcommand{\ba}{\begin{eqnarray}}
\newcommand{\ea}{\end{eqnarray}}

\newcommand{\beq}{\begin{equation}}
\newcommand{\eeq}{\end{equation}}
\newcommand{\beqa}{\begin{eqnarray}}
\newcommand{\eeqa}{\end{eqnarray}}

\begin{document}

\title{Slowly rotating black holes with exact Killing tensor symmetries}

\author{Finnian Gray}
\email{fgray@perimeterinstitute.ca}
\affiliation{Perimeter Institute, 31 Caroline Street North, Waterloo, ON, N2L 2Y5, Canada}
\affiliation{Department of Physics and Astronomy, University of Waterloo,
Waterloo, Ontario, Canada, N2L 3G1}

\author{David Kubiz\v n\'ak}
\email{dkubiznak@perimeterinstitute.ca}
\affiliation{Perimeter Institute, 31 Caroline Street North, Waterloo, ON, N2L 2Y5, Canada}
\affiliation{Department of Physics and Astronomy, University of Waterloo,
Waterloo, Ontario, Canada, N2L 3G1}

\date{February 15, 2022}

\begin{abstract}
We present a novel family of slowly rotating black hole solutions in four, and higher dimensions, that extend the well known Lense--Thirring spacetime and solve the field equations to linear order in rotation parameter. As ``exact metrics'' in their own right, the new (non-vacuum) spacetimes feature the following two remarkable properties: i) near the black hole horizon they can be cast in the, manifestly regular, Painlev\'e--Gullstrand form and ii) they admit exact Killing tensor symmetries. We show these symmetries are inherited from the principal Killing--Yano tensor of the exact rotating black hole geometry in the slow rotation limit. This provides a missing link as to how the exact hidden symmetries emerge as rotation is switched on. Remarkably, in higher dimensions the novel generalized Lense--Thirring spacetimes feature a rapidly  growing number of exact irreducible rank-2, as well as higher-rank, Killing tensors -- giving a first example of a physical spacetime with more hidden than explicit symmetries.
\end{abstract}

\maketitle

\section{Introduction}

Within the framework of Hamiltonian dynamics one can distinguish two kinds of symmetries: explicit symmetries, that induce a non-trivial action on the configuration space, and,  dynamical symmetries, which are the genuine symmetries of the phase space. Since the latter ``remain hidden'' on the configuration space, they are sometimes referred to as {\em hidden symmetries}.

Perhaps the most familiar example of a hidden symmetry in the context of general relativity  is associated with a {\em Killing tensor} \cite{Walker:1970un}. This is a symmetric rank-$p$ tensor $K^{\alpha_1\dots \alpha_p}=K^{(\alpha_1\dots \alpha_p)}$ obeying the following Killing tensor equation:
\be\label{KT}
\nabla^{(\beta}K^{\alpha_1\dots \alpha_p)}=0\,,
\ee
which for $p=1$ yields a Killing vector. Eq.~\eqref{KT} represents an overdetermined partial differential equation that imposes severe restrictions on the background metric, e.g. \cite{Houri:2017tlk}. Once present in a given spacetime, Killing tensors give rise to monomial integrals of motion for geodesic trajectories -- the most famous example being {\em Carter's constant} in the Kerr geometry \cite{Carter:1968rr}. Killing tensors may also give rise to symmetry operators of the scalar wave equation and underlie its separability \cite{Carter:1977pq}.

Killing tensors form a subalgebra of the full algebra of symmetric tensor fields 
under an operation called the {\em Schouten--Nijenhuis (SN) bracket} \cite{schouten1940uber,nijenhuis1955jacobi}. That is, given two Killing tensors $A^{\alpha_1\dots \alpha_p}$ and $B^{\beta_1\dots \beta_q}$, their SN bracket
\ba\label{SN}
[A,B]_{\mbox{\tiny SN}}^{\alpha_1\dots \alpha_{p+q-1}}&=&
pA^{\gamma (\alpha_1\dots \alpha_{p-1}}\nabla_\gamma B^{\alpha_p\dots \alpha_{p+q-1})}\nonumber\\
&&\!\!\!\!\!\!\!\!\!\!-qB^{\gamma(\alpha_1\dots \alpha_{q-1}}\nabla_\gamma A^{\alpha_q \dots \alpha_{q+p-1})}\, 
\ea
yields another Killing tensor. In terms of these brackets, the Killing tensor equation \eqref{KT} is conveniently expressed as $[K, g]_{\mbox{\tiny SN}}^{\alpha_1\dots \alpha_p}=0$\,.

A trivial example of a Killing tensor is the metric itself, another is obtained by taking a symmetrized product of Killing vectors. If a Killing tensor cannot be decomposed into a linear combination of products of lower rank Killing tensors, it is called {\em irreducible}. Reducible Killing tensors are trivial in the sense that they generate no new conserved quantities and can typically be excluded from further considerations. While spacetimes with exact irreducible Killing tensors are quite rare, the pivotal examples include the Kerr family of black hole spacetimes in all dimensions \cite{Frolov:2003en, Frolov:2017kze} as well as various supergravity solutions,  e.g. \cite{Chong:2005hr, Chow:2008fe}.

Another, perhaps even more intricate,  example of a hidden symmetry is that of {\em Killing--Yano tensors} \cite{yano1952some} which are in some sense the square root of Killing--tensors. In particular, of special importance for black hole physics is the principal Killing--Yano tensor \cite{floyd1973dynamics, Frolov:2006dqt}, which is a 2-form $h_{\alpha\beta}$ obeying the following equations:
\be\label{eq:principalh}
\nabla_\gamma h_{\alpha\beta}=2g_{\gamma[\beta}\xi_{\alpha]}\,,\quad \xi_\alpha=-\frac{1}{3}\nabla_\gamma h^\gamma{}_\alpha\,. 
\ee
It turns out that starting with one such object, one may be able to generate a whole tower of Killing tensors. The first of these (and the only one in four dimensions) is given by 
\be\label{K}
K_{\alpha\beta}=\bigl((*h)\cdot (*h)\bigr)_{\alpha\beta}=Q_{\alpha\beta}-\frac{1}{2}g_{\alpha\beta}Q^\gamma{}_\gamma\,,
\ee 
where $Q_{\alpha\beta}=h_{\alpha\gamma}h_\beta{}^\gamma$, and we have defined $(\omega \cdot \omega)_{\alpha\beta}=\frac{1}{p!}\omega_{\alpha\gamma_1\dots \gamma_p}\omega_\beta{}^{\gamma_1\dots \gamma_p}$ for any $(p+1)$-form $\omega$. Notably,  this construction applies to the Killing tensor of the Kerr geometry \cite{floyd1973dynamics} and its higher-dimensional generalizations \cite{Frolov:2017kze}, where it guarantees the complete integrability of geodesic motion \cite{Carter:1968rr, Frolov:2003en, Page:2006ka}.

It is the aim of the present paper to construct a new class of solutions with exact Killing tensors. Namely, we pick up the threads on the recent observation \cite{Baines:2021qaw} that an appropriately modified {\em Lense--Thirring spacetime} \cite{lense1918influence}, which describes a field of a slowly rotating body, admits the exact Killing tensor. We show that this result can be extended to the whole family of (possibly charged) {\em generalized Lense--Thirring spacetimes} with a cosmological constant in four and higher dimensions. 

These spacetimes are ``derived'' from the corresponding exact black hole solutions (seeds) in the slow rotation approximation. As such they inherit the approximate hidden symmetries of the exact seed solutions. Remarkably, when these metrics are ``appropriately modified'', the approximate Killing tensor symmetry becomes exact. In addition, the obtained spacetimes are regular on the black hole horizon and close to its vicinity can be cast in the {\em Painlev\'e--Gullstrand (PG) form}, see \cite{Martel:2000rn, Faraoni:2020ehi}.

\section{Exact Kerr--Newmann-AdS solution and its slow rotation expansion}\label{Sec2}

To start our discussion, let us recapitulate the exact Einstein--Maxwell-$\Lambda$ solution for a rotating black hole -- known as the Kerr--Newmann-AdS metric 
\cite{Carter:1968ks}, which we write in the ``standard Boyer--Lindquist form''~\cite{Hawking:1998kw}:
\ba\label{KNADS2}
ds^2&=&-\frac{\Delta}{\Sigma}\left[dt-\frac{a\sin^2\!\theta}{\Xi}d\phi\right]^2
+\frac{\Sigma}{\Delta} dr^2+\frac{\Sigma}{S}d\theta^2\nonumber\\
&&+\frac{S\sin^2\!\theta}{\Sigma}\left[a dt-\frac{r^2+a^2}{\Xi}d\phi\right]^2\,,\nonumber\\
{A}&=&-\frac{qr}{\Sigma}\left(dt-\frac{a\sin^2\!\theta}{\Xi}d\phi\right)\,,
\ea
where $A$ is the vector potential and $F=dA$ the corresponding field strength, $a$ is the rotation parameter,
\ba\label{KerrSigmaa}
\Sigma&=&r^2+a^2\cos^2\!\theta\,,\quad \Xi=1-\frac{a^2}{\ell^2}\,,
\quad S=1-\frac{a^2}{\ell^2}\cos^2\!\theta\,,\nonumber\\
\Delta&=&(r^2+a^2)\Bigl(1+\frac{r^2}{\ell^2}\Bigr)-2mr+q^2\,,
\ea
$m$ and $q$ are the mass and charge parameters, and $\ell$ is the AdS radius. One can check that the geometry at the horizon, located at $r=r_+$, given by the largest root $\Delta(r_+)$ is regular, and in particular, the Kretschmann scalar,
\be\label{I}
I=R_{\alpha\beta\gamma\delta} R^{\alpha\beta\gamma\delta}\,,
\ee
is smooth at $r=r_+$. As written, the solution rotates at infinity -- this rotation can be removed by ``going to the non-rotating frame'': $d\phi\to d\phi-a/\ell^2 dt$.

The metric is the algebraically special type D and admits a fundamental hidden symmetry, encoded in the principal Killing--Yano tensor $h$, which obeys~\eqref{eq:principalh}, 
and is explicitly given by  
$h=db$\,, 
where 
\be
2b=r^2\bigl(dt-a\sin^2\!\theta\frac{d\phi}{\Xi}\bigr)-a^2\cos^2\!\theta \bigl(dt-\frac{ad\phi}{\Xi}\bigr)\,. 
\ee
The corresponding irreducible Killing tensor, constructed from $h$ according to \eqref{K}, reads:
\ba
K&=&\frac{a^2\cos^2\!\theta}{\Delta \Sigma}\Bigl((r^2+a^2)\partial_t+a\Xi\partial_\phi\Bigr)^2-\frac{a^2\cos^2\!\theta \Delta}{\Sigma}(\partial_r)^2\nonumber\\
&&+\frac{r^2}{\Sigma S \sin^2\!\theta}(a\sin^2\!\theta \partial_t +\Xi \partial_\phi)^2+\frac{Sr^2}{\Sigma} (\partial_\theta)^2\,.
\ea
Together with  the two explicit symmetries, $\partial_t$ and $\partial_\phi$, it guarantees the complete integrability of geodesic motion in these spacetimes \cite{Carter:1968ks, carter1968new}.

Let us now perform the linear in $a$ expansion to the above exact solution. This yields the following approximate to $O(a)$ solution of the Einstein--Maxwell-$\Lambda$ equations:
\ba
ds^2&=&-f dt^2+\frac{dr^2}{f}+2a\sin^2\!\theta(f-1)dt  d\phi\,,\nonumber \\
&&+r^2\sin^2\!\theta d\phi^2 +r^2 d\theta^2+O(a^2)\,,\nonumber\\
A&=&-\frac{q}{r}\bigr(dt-a\sin^2\!\theta d\phi\bigr)+O(a^2)\,,\label{Oa}
\ea
where  
\be\label{f}
f=1-\frac{2M}{r}+\frac{q^2}{r^2}+\frac{r^2}{\ell^2}\,. 
\ee

Of course, the spacetime inherits the hidden symmetries of the full solution to the linear order in $a$, given by 
\ba
2b&=&{r^2}dt-{ar^2\sin^2\!\theta}d\phi +O(a^2)\,,\label{ba}\\
K&=&2a \partial_t \partial_\phi +\frac{1}{\sin^2\!\theta}(\partial_\phi)^2 +(\partial_\theta)^2+O(a^2)\,.\label{Ka}
\ea
Note that since the metric is stationary and axisymmetric, the first term in \eqref{Ka} is just a product of Killing vectors and can be excluded.

A tempting possibility is to truncate the $O(a^2)$ terms in \eqref{Oa}, and treat the resultant fields as ``exact'' (not necessarily a solution of the field equations).
However, the spacetime has several ``drawbacks''. Namely, as exact metric, it is singular on what would be the black hole horizon $f=0$, noting for example that the Kretschmann scalar \eqref{I} diverges there at $O(a^2)$. Second, both (truncated to $O(a)$)  hidden symmetries \eqref{ba} and \eqref{Ka} remain only approximate. Finally, the metric cannot be cast in the PG form \cite{Baines:2020unr}.

\section{Generalized Lense--Thirring metric}

To fix the above ``drawbacks'', let us instead consider the following modification of the above slowly rotating solution:  
\ba\label{Lense}
ds^2&=&-f dt^2+\frac{dr^2}{f}+r^2\sin^2\!\theta\Bigl(d\phi \!+\!\frac{a(f-1)}{r^2}dt\Bigr)^2\!\!+\!r^2 d\theta^2\,,\nonumber \\
A&=&-\frac{q}{r}\Bigl(dt-a\sin^2\!\theta \bigl[d\phi+\frac{a}{fr^2}dr
\bigr]\Bigr)\,,
\ea
with metric function $f$ given by Eq.~\eqref{f}.   
In what follows, we shall call it the {\em generalized Lense--Thirring solution}, c.f. \cite{Baines:2020unr}. Formally, it can be obtained by ``completing the square'' in the truncated solution \eqref{Oa} (together with an appropriate modification of the vector potential $A$ to achieve regularity of the electromagnetic invariants on the horizon).
As such, it still solves the Einstein--Maxwell-$\Lambda$ system to $O(a)$, as well as admitting the approximate hidden symmetries \eqref{ba} and \eqref{Ka}.

However, when understood as an exact (filled with matter) spacetime,\footnote{One should be a bit cautious about the physicality of the extra matter supporting the (modified) Lense--Thirring spacetimes.  In particular, when $Q=0=\Lambda$, the tetrad component of the Einstein tensor is $G^{\hat 0 \hat 0}=-\frac{1}{2}R=-(3aM\sin\theta/r^4)^2$, yielding a negative energy density \cite{Baines:2020unr}. This is partly emended in the presence of the additional ordinary matter -- for example in our case the EM field dominates near infinity and guarantees there positive energy density. However, we expect that the generic Lense--Thirring spacetimes may need to be  (at least partly) supported by exotic matter violating the standard energy conditions. 
}   
it is a much better approximation for a slowly rotating black hole than the above truncated solution since it is regular on the horizon -- the curvature scalars, such as $I$ \eqref{I}, no longer diverge at $f=0$ and the metric can be cast (at least in the vicinity of the horizon) in the manifestly regular PG form, see appendix~\ref{AppendixA} for more details. It also has an ergosphere and will feature superradiant phenomena, e.g. \cite{Brito:2015oca}. Most remarkably, the generalized Lense--Thirring spacetime \eqref{Lense} falls into a class of the Benenti and Francaviglia metrics~\cite{Benenti:1979} (see also \cite{Papadopoulos:2020kxu} for some recent applications). This means that not only does the metric posses an exact Killing tensor, the separability of the scalar wave equation and integrability of the geodesics are guaranteed.\footnote{In particular, one can check that Carter's criterion~\cite{Carter:1977pq} $\nabla_\alpha(k_\gamma{}^{[\alpha}R^{\beta]\gamma})$=0 is satisfied. Interestingly, one can also check that the criteria required for separability of the conformally coupled scalar equation are not satisfied, even though this equation does separate for Kerr--NUT--AdS spacetimes~\cite{Gray:2021wzf}.}. 

The corresponding exact Killing tensor is given by 
\be\label{Kexact}
K=\frac{1}{\sin^2\!\theta}(\partial_\phi)^2+(\partial_\theta)^2\,,
\ee
and can be understood as a slow rotation (truncated) version of the approximate Killing tensor \eqref{Ka}. Interestingly, this Killing tensor can be written in the following suggestive form:
\be
K=L_x^2+L_y^2+L_z^2\,,  
\ee
where $L_z=\partial_\phi$ is a Killing vector of \eqref{Lense} and vectors $L_x$ and $L_y$ are given by  
\ba
L_x&=&\cot \theta \cos\phi \partial_\phi+\sin \phi \partial_\theta\,,\nonumber\\
L_y&=&-\cot \theta \sin\phi \partial_\phi+\cos \phi \partial_\theta\,,
\ea
which upon recovering the spherical symmetry ($a\to 0$) would be the remaining two $SO(3)$ Killing vectors. Since $L_z$ and $\partial_t$ are the only two Killing vectors present in the spacetime \eqref{Lense}, it can be checked that the above Killing tensor is irreducible.

Moreover, we may define the following 2-form:
\be
h^{(0)}=d b^{(0)}\,\quad 2b^{(0)}=r^2 dt\,, 
\ee
obtained by the $a\to 0$ limit of the 2-form \eqref{ba}. While this is not a principal tensor even to the linear order in $a$, it yields the above exact Killing tensor \eqref{Kexact} via the formula \eqref{K} (with $h\to h^{(0)}$). 
We also note that 
\be
\xi^{(0)}=-\frac{1}{3}\nabla \cdot h^{(0)} =\partial_t+a\Bigl(\frac{q^2}{3r^4}-\frac{1}{\ell^2}\Bigr)\partial_\phi\,, 
\ee
which is an \emph{exact} Killing vector when $q=0$.

Let us finally mention that the generalized Lense--Thirring spacetime \eqref{Lense} is, contrary to the exact solution \eqref{KNADS2}, algebraically general and describes a slowly rotating charged black hole (or rotating body) that can be assigned the following asymptotic charges:
\be
M=m\,,\quad J=ma\,,\quad Q=q\Bigl(1+\frac{2a^2}{3\ell^2}\Bigr)\,,
\ee
and is surrounded by (charged) matter. To linear order in $a$, the corresponding first law of black hole thermodynamics coincides with that of the spherical charged AdS black hole, e.g. \cite{Kubiznak:2012wp}.

\section{Higher-dimensional Lense--Thirring spacetimes}\label{Sec4}

The above construction becomes even more remarkable in higher dimensions. To illustrate this, let us start from the full Kerr-AdS metric in $d$ spacetime dimensions 
\cite{Gibbons:2004js}:
\ba\label{HDKerr}
ds^2&=&-W(1+r^2/\ell^2)dt^2+\frac{2M}{U}\Bigl(Wdt+\sum_{i=1}^m \frac{a_i \mu_i^2 d\phi_i}{\Xi_i}\Bigr)^2\nonumber\\
&&+\sum_{i=1}^m\frac{r^2+a_i^2}{\Xi_i}\bigl(\mu_i^2d \phi_i^2+d\mu_i^2)
+\frac{Udr^2}{V-2M}+\epsilon r^2 d\nu^2\nonumber\\
&&+\frac{1}{W(l^2-r^2)}\Bigl(\sum_{i=1}^m \frac{r^2+a_i^2}{\Xi_i}\mu_i d\mu_i+\epsilon r^2 \nu d\nu\Bigr)^2\,, 
\ea 
where 
\ba
 W&=&\sum_{i=1}^m \frac{\mu_i^2}{\Xi_i}+\epsilon \nu^2\,,\quad V=r^{\epsilon-2}(1+r^2/\ell^2)\prod_{i=1}^m(r^2+a_i^2)\,,\nonumber\\
U&=&\frac{V}{1+r^2/\ell^2}\Bigl(1-\sum_{i=1}^m\frac{a_i^2\mu_i^2}{r^2+a_i^2}\Bigr)\,,\quad \Xi_i=1-\frac{a_i^2}{\ell^2}\,. 
\ea
Here, $\epsilon=1, 0$ for even, odd dimensions, $m=\bigl[\frac{d-1}{2}\bigr]$ (where $[A]$ denotes the whole part of $A$), and the coordinates $\mu_i$ and $\nu$ obey a constraint 
\be 
\sum_{i=1}^m \mu_i^2+\epsilon \nu^2=1\,.
\ee 

The metric admits \cite{Frolov:2006dqt, Frolov:2017kze} a principal Killing--Yano tensor, $h=db$,  
\be\label{hHD}
2b=\Bigl(r^2+\sum_{\mu=1}^m a_i^2\mu_i^2(1+\frac{r^2+a_i^2}{\ell^2 \Xi_i}\Bigr)dt
+\sum_{i=1}^m a_i\mu_i^2 \frac{r^2+a_i^2}{\Xi_i}d\phi_i\,, 
\ee
which generates the towers of explicit and hidden symmetries, see \cite{Frolov:2017kze}. 

By repeating the procedure above, we arrive at the following slowly rotating generalized Lense--Thirring solution (written now in non-rotating at infinity coordinates): 
\ba\label{LTHD}
ds^2&=&-fdt^2+\frac{dr^2}{f}+r^2\sum_{i=1}^m \mu_i^2\Bigl(d\phi_i+\frac{2Ma_i}{r^{2m+\epsilon}} dt\Bigr)^2\nonumber\\
\!\!\!&+&\!r^2(\sum_{i=1}^m\!d\mu_i^2\!+\!\epsilon d\nu^2)\,,\ f\!=\!1-\frac{2M}{r^{2m-2+\epsilon}}+\frac{r^2}{\ell^2}\,.\ 
\ea
As before, the metric is regular on the horizon, $f=0$, near its vicinity admits the PG form (see appendix~\ref{AppendixA}), and inherits the following approximate principal Killing--Yano tensor: 
\be
2b=r^2 dt+r^2\sum_{i=1}^m a_i \mu_i^2 d\phi_i\,. 
\ee

Surprisingly, in addition, we have a fast growing (with number of dimensions) tower of exact Killing tensors. Explicitly,
let us define the set $S=\{1,..,m\}$ and let $I\in P(S)$ where $P(S)$ is the power set of $S$, then we have the following objects:
\begin{align}
	2b^{(I)}&\equiv r^2(dt+\sum_{i\in I}a_i\mu_i^2d\phi_i)\,,\quad\,h^{(I)}\equiv db^{(I)}\,,\\
	f^{(I)}&\equiv \frac{1}{(|I|+1)!}*\big(\underbrace{h^{(I)}\wedge \dots \wedge h^{(I)}}_{|I|+1\ \mbox{\tiny times}}\big)\,,
\end{align}
where $|I|$ denotes the size of the set $I$. These generate the following exact rank-2 Killing  tensors:
\begin{align}
K^{(I)}_{\mu\nu}&=\big(\prod_{i\in I} a_i\big)^{-2}\,(f^{(I)}\cdot f^{(I)})_{\mu\nu}\,.
\end{align}
Note that, this construction ``coincides'' with the one for the full Kerr-AdS geometry \cite{Frolov:2017kze}, replacing the principal Killing--Yano tensor $h$ with its appropriate limits $h^{(I)}$ at the relevant order of the small rotation parameters expansion.

Of course, in a given dimension, not all of these are non-trivial. In fact, it is only $K^{(\emptyset)}$ which exists in all dimensions $d\geq 4$, and is given by our familiar formula \eqref{K} (with $h\to h^{(\emptyset)}$).
Explicitly, these Killing tensors take the following simple form:
\begin{align}
K^{(I)}&=\!\!\!\!\!\sum\limits_{i\not\in I}^{m-1+\epsilon}\!\bigg[\bigr(1-\mu_i^2-\!\sum_{j\in I}\mu_j^2\bigr)(\partial_{\mu_i})^2 -2\!\!\!\!\sum_{j\not\in I\cup\{i\}}\!\!\!\! \mu_i\mu_j\,\partial_{\mu_i}\partial_{\mu_j} \bigg] \nonumber\\
&+\sum\limits_{i\not\in I}^{m}\bigg[\frac{1-\sum_{j\in I}\mu^2_j}{\mu_i^2}(\partial_{\phi_i})^2\bigg] \,.
\end{align}
It turns out, however, that 
 $\sum_{i=0}^{m-3}{m \choose i}$ of these are reducible, leaving 
\be
k=\sum_{i=0}^{m-2+\epsilon}{m \choose i}-\sum_{i=0}^{m-3} {m \choose i} =\frac{1}{2} m (m-1+2\epsilon)
\ee
irreducible rank-2 Killing tensors in $d$ dimensions. Contrary to the exact Kerr-AdS spacetimes whose number of rank-2 Killing tensors grows linearly with number of spacetime dimensions, in our case the growth is {\em quadratic}. For example, already in $d=8$ we have (for distinct rotation parameters) 6 irreducible rank-2 Killing tensors and only $m+1=4$ independent Killing vectors -- that is the number of hidden symmetries exceeds the number of the explicit ones (more so once we also count higher rank Killing tensors obtained by various combinations of SN brackets -- see below). However, this is still much smaller than the maximum possible number of rank-2 Killing tensors in a given dimension $d$, which for rank-$p$ Killing tensor $(p\geq 1)$ reads, e.g. \cite{Houri:2017tlk}: 
\be 
k_{\max}=\frac{1}{d}\Bigl(\begin{matrix}
d+p\\
p+1
\end{matrix}\Bigr) 
\Bigl(\begin{matrix}
d+p-1\\
p
\end{matrix}\Bigr) \,,
\ee 
and for $d=8$ and $p=2$ gives $k_{\max}=540$.

In addition to the above rank-2 Killing tensors, one can also generate (potentially irreducible) higher-rank Killing tensors via the SN brackets \eqref{SN}. 
Our construction thus provides a physically well motivated example in the long-standing search for spacetimes with higher-rank Killing tensors~\cite{Brink:2008xy, Gibbons:2011hg,Cariglia:2015fva}. Of course, these objects are, at the same time, approximate (to linear order in rotation parameters) higher-rank Killing tensors for the full Kerr-AdS geometry \eqref{HDKerr}. 
We have verified up to $d=13$ that the SN bracket of any two Killing tensors vanishes if the intersection of the two labels equals the first. That is,
\be
[K^{(I_1)}, K^{(I_2)}]_{\mbox{\tiny SN}}=0\,,
\ee
if $I_1\cap I_2=I_1$. Otherwise, a new Killing tensor is generated. 
In particular, in $d$ dimensions  this implies (taking into account explicit symmetries and the metric as well) the existence of $d$ mutually commuting 
symmetry objects -- a necessary requirement for complete (and again exact) integrability of geodesic motion in the spacetime \eqref{LTHD}.
We expect this to remain true also in higher dimensions. 

Finally we close by illustrating the above construction in $d=6$ dimensions. In this case $k=3$ and we have the following irreducible rank-2 Killing tensors:
\ba
K^{(\emptyset)}&=&\frac{1}{\mu_1^2}(\partial_{\phi_1})^2+\frac{1}{\mu_2^2}(\partial_{\phi_2})^2
+(1-\mu_1^2)(\partial_{\mu_1})^2\nonumber\\
&&-2\mu_1\mu_2(\partial_{\mu_1})(\partial_{\mu_2})+(1-\mu_2^2)(\partial_{\mu_2})^2\,,\nonumber\\
K^{(1)}&=&\frac{1-\mu_1^2}{\mu_2^2}(\partial_{\phi_2})^2-(1-\mu_1^2-\mu_2^2)(\partial_{\mu_2})^2\,,\\
K^{(2)}&=&\frac{1-\mu_2^2}{\mu_1^2}(\partial_{\phi_1})^2-(1-\mu_1^2-\mu_2^2)(\partial_{\mu_1})^2\,. \nonumber
\ea
Their SN brackets are 
\be
[K^{(\emptyset)},K^{(1)}]_{\mbox{\tiny SN}}=0=[K^{(\emptyset)},K^{(2)}]_{\mbox{\tiny SN}}\,,\  M=[K^{(1)},K^{(2)}]_{\mbox{\tiny SN}}\,, 
\ee
where $M$ is the new rank-3 Killing tensor with the following components:
\ba
M^{\mu_1\mu_2\mu_2}&=&-\frac{4\mu_1(\mu_1^2+\mu_2^2-1)}{3}=\mu_2^2M^{\phi_2\phi_2\mu_1}\,,\nonumber\\ 
M^{\mu_1\mu_1\mu_2}&=&\frac{4\mu_2(\mu_1^2+\mu_2^2-1)}{3}=\mu_1^2M^{\phi_1\phi_1\mu_2}\,.
\ea
Provided no additional irreducible rank-2 Killing tensors exist in this spacetime, $M$ is also irreducible.  
This tensor further generates rank-4 Killing tensors via SN brackets with $K^{(1)}$ and $K^{(2)}$, and so on.

\section{Conclusions}\label{Sec5}
Starting in four dimensions, we have seen how a ``small modification'' of the linear in $a$ expansion of the exact Kerr--Newmann-AdS black hole solution gives rise to an, in many ways, preferred slowly rotating geometry.  This {\em generalized Lense--Thirring spacetime}, is (when taken as an exact metric) manifestly regular on the black hole horizon and admits an exact Killing tensor.

This observation fills an important gap in understanding as to how the exact hidden symmetries of the full Kerr--Newmann-AdS geometry emerge as the rotation is switched on. While the non-rotating (spherical) solution admits an exact principal Killing--Yano and Killing tensor, these are trivial, the latter being reducible -- given by a product of Killing vectors derived from the rotational symmetry (and possibly time independence). Adding a small rotation to $O(a)$ breaks the full rotational symmetry and the approximate hidden symmetries become non-trivial. Remarkably a simple modification of the metric at $O(a^2)$ yields a spacetime which in 4 dimensions is of the Benenti  and Francaviglia class of spacetimes \cite{Benenti:1979} in which separability of the Klein-Gordon and Hamilton--Jacobi equations is guaranteed. The exact irreducible Killing tensor can be understood as a (truncated) version of the approximate Killing tensor generated from the approximate principal Killing--Yano tensor.  Both these hidden symmetries become exact, when the full Kerr--Newmann-AdS solution is considered.

Naturally, a similar construction also works in higher dimensions, which we have explicitly demonstrated for Kerr-AdS spacetimes in all dimensions, however the structure is much richer. The corresponding generalized Lense--Thirring spacetimes admit a rapidly growing  tower of exact rank-2 and higher-rank Killing tensors, that is a ``slow rotation seed'' of the associated (much smaller) tower of rank-2 Killing tensors for the full Kerr-AdS geometry. Although the higher-dimensional Lense--Thiring spacetime \eqref{LTHD} is not explicitly in the Benenti--Francaviglia form, we have a tower of Killing tensors growing faster than the number of Killing vectors -- providing a first example of a physically interesting spacetime with larger number of hidden symmetries than the explicit ones.

We expect this construction to be quite general and apply to many other rotating black hole spacetimes with hidden symmetries (e.g. \cite{Chong:2005hr, Chow:2008fe}). In particular, note that while in our paper we have focused on negative $\Lambda$, the same construction would also work for $\Lambda$ positive.
It remains an interesting open question whether similar construction would also work for higher order expansions in rotation parameters, providing thus even a more complete link between the generalized Lense--Thirring spacetimes and the exact black hole solutions.

\section*{Acknowledgements}
F.G. acknowledges support from the Natural Sciences and Engineering Research Council of Canada (NSERC) via a Vanier Canada Graduate Scholarship. This work was supported by the Perimeter Institute for Theoretical Physics and by NSERC. Research at Perimeter Institute is supported in part by the Government of Canada through the Department of Innovation, Science and Economic Development Canada and by the Province of Ontario through the Ministry of Colleges and Universities. 
Perimeter Institute and the University of Waterloo are situated on the Haldimand Tract, land that was promised to the Haudenosaunee of the Six Nations of the Grand River, and is within the territory of the Neutral, Anishnawbe, and Haudenosaunee peoples.


\providecommand{\href}[2]{#2}\begingroup\raggedright\endgroup

\appendix

\section{Painlev\'e--Gullstrand form}\label{AppendixA}

In what follows we shall demonstrate that the generalized Lense--Thirring spacetimes can, at least in the vicinity of the black hole horizon, be cast in the PG form  \cite{Martel:2000rn, Faraoni:2020ehi}. As we shall see this can be formally achieved by a simple coordinate transformation, c.f. \cite{Baines:2020unr,Baines:2021qaw}. 

Let us start in four dimensions, having the solution \eqref{Lense}. The PG coordinates are traditionally associated with a free-falling observer starting from rest at infinity and moving (at infinity) radially inward. However, since our metrics have possibly AdS asymptotics, this is no longer achievable (due to the AdS ``attraction'' timelike geodesics do not reach asymptotic infinity). Formally, however, one can consider radially infalling observers starting from rest at a finite radius from the black hole, determined for simplicity by $f(r_0)=1$ -- restricting to regions with $f\leq 1$. (For asymptotically flat spacetimes $r_0$ approaches infinity.) The corresponding  
4-velocity then reads:
\be
 u=dt+\frac{\sqrt{1-f}}{f}dr\,.
\ee
Setting the latter equal to $dT$, where $T$ is the proper time of the observer, we arrive at the following coordinate transformation:
\be
dt=dT-\frac{\sqrt{1-f}}{f}dr\,, 
\ee 
upon which the solution \eqref{Lense} can be written as 
\ba
ds^2&=&-dT^2+(dr+\sqrt{1-f}dT)^2+r^2d\theta^2\nonumber\\
&+&r^2\sin^2\!\theta \Bigl(d\phi+\frac{a(f-1)}{r^2}dT-\frac{a(f-1)\sqrt{1-f}}{r^2f}dr\Bigr)^2\,,\nonumber\\
A&=&-\frac{q}{r}\Bigl(dT-a\sin^2\!\theta \bigl[d\phi+\frac{a}{fr^2}dr
\bigr]\Bigr)\,. 
\ea
Here we have dropped the pure gauge term proportional to $dr$. Formally, one can bring the metric in the PG form by setting 
\be
d\phi=d\Phi +\frac{a(f-1)\sqrt{1-f}}{r^2f}dr\,,
\ee
upon which we recover
\ba
ds^2&=&-dT^2+(dr+\sqrt{1-f}dT)^2+r^2d\theta^2\nonumber\\
&&+r^2\sin^2\!\theta \Bigl(d\Phi+\frac{a(f-1)}{r^2}dT\Bigr)^2\,,\\
A&=&-\frac{q}{r}\Bigl(dT-a\sin^2\!\theta \Bigl[d\Phi+\frac{a(1+\sqrt{1-f}(f-1))}{fr^2}dr\Bigr]\Bigr)\,. \nonumber
\ea
Obviously, the spatial hypersurfaces $T=$const. are all intrinsically flat, and the metric as well as the vector potential are manifestly non-singular for $f=0$. The fact that the Lense--Thirring spacetimes can be brought into the PG form by a coordinate transformation shows that the Killing tensor discovered in \cite{Baines:2021qaw} coincides with the one studied in this paper for vanishing cosmological constant and $q=0$. At the same time it shows that the metrics (2.3) and (2.4) in \cite{Baines:2020unr} are at least locally diffeomorphic.

Similarly, the higher-dimensional Lense--Thirring spacetimes \eqref{LTHD} can be cast in the higher-dimensional version of the PG form close to the horizon, by the following change of coordinates:
\ba
dt&=&dT-\frac{\sqrt{1-f}}{f}dr\,,\nonumber\\
d\phi_i&=&d\Phi_i+\frac{2Ma_i\sqrt{1-f}}{f r^{2m+\epsilon}}dr\,, 
\ea
upon which the metric takes the following PG form: 
\ba\label{LTHDPG}
ds^2&=&-dT^2+(dr+\sqrt{1-f}dT)^2
+r^2(\sum_{i=1}^md\mu_i^2+\epsilon d\nu^2)\nonumber\\
&&+r^2\sum_{i=1}^m \mu_i^2\Bigl(d\Phi_i+\frac{2Ma_i}{r^{2m+\epsilon}} dT\Bigr)^2\,,\nonumber\\
f&=&1-\frac{2M}{r^{2m-2+\epsilon}}+\frac{r^2}{\ell^2}\,, 
\ea
which is non-singular on the horizon $f=0$ and whose $T=$const. slices are manifestly flat.

\end{document}